# Magnetic-field modulation of topological electronic states and emergent magneto-transport in a magnetic Weyl semimetal

Jianlei Shen,[1,2,*] Jiacheng Gao,[1,3,*] Changjiang Yi,[1,*] Meng Li,[1,3,*] Qingqi Zeng,[1] Shen Zhang,[1,3] Jinying Yang,[1,3] Binbin Wang,[1] Min Zhou,[4] Rongjin Huang,[4] Hongxiang Wei,[1] Haitao Yang,[1] Youguo Shi,[1,5] Geng Li,[1,5,†] Zhijun Wang,[1,3,†] Enke Liu,[1,3,5†] Xiaohong Xu,[2] Hongjun Gao,[1,5] Baogen Shen[1,3,6,7]

[1]Beijing National Laboratory for Condensed Matter Physics, Institute of Physics, Chinese Academy of Sciences, Beijing 100190, China

[2]Research Institute of Materials Science of Shanxi Normal University, Taiyuan 030000, China

[3]School of Physical Sciences, University of Chinese Academy of Sciences, Beijing 101408, China

[4]Key Laboratory of Cryogenics, Technical Institute of Physics and Chemistry, Chinese Academy of Sciences, Beijing 100190, China.

[5]Songshan Lake Materials Laboratory, Dongguan 523808, China

[6]Ningbo Institute of Materials Technology & Engineering, Chinese Academy of Sciences, Ningbo, Zhejiang, 315201, China

[7]Ganjiang Innovation Academy, Chinese Academy of Sciences, Ganzhou, Jiangxi, 341000, China

**The modulation of topological electronic states by an external magnetic field is highly desired for condensed matter physics. Schemes to achieve this have been proposed theoretically, but few can be realized experimentally. Here, combining transverse transport, theoretical calculations, and scanning tunneling microscopy/spectroscopy (STM/S) investigations, we provide an observation that the topological electronic states, accompanied by an emergent magneto-transport phenomenon, were modulated by applying magnetic fields through induced non-collinear magnetism in the magnetic Weyl semimetal $EuB_6$. A giant unconventional anomalous Hall effect (UAHE) is found during the magnetisation re-orientation from easy axis to hard ones in magnetic fields, with a UAHE peak around the low field of 5 kOe. Under the reasonable spin-canting effect, the folding of the topological anti-crossing bands occurs, generating a strong Berry curvature that accounts for the observed UAHE. Field-dependent STM/S reveals a highly synchronous evolution of electronic density of states, with a d*I*/d*V* peak around the same field of 5 kOe, which provides evidence to the folded bands and excited UAHE by external magnetic fields. This finding elucidates the connection between the real-space non-collinear magnetism and the *k*-space topological electronic states and establishes a novel manner to engineer the magneto-transport behaviours of correlated electrons for future topological spintronics.**

[*]These authors contributed equally to this work.

[†] Corresponding authors, ekliu@iphy.ac.cn (E. L.), wzj@iphy.ac.cn (Z. W.), gengli.iop@iphy.ac.cn (G. L.)

From the perspective of applications of spintronics and future quantum devices, it is highly desirable to modulate the electronic structures using external fields, and further to control the electronic transport behaviour. Recently, materials with topologically nontrivial electronic states have attracted much attention [1-4]. In particular, magnetic topological materials [5-9], where the transport behaviour, such as the intrinsic anomalous Hall effect (AHE) [5,6,10], anomalous Nernst effect [11,12], and quantum anomalous Hall effect [9,13], are dominated by subtle topological electronic bands near the Fermi level. Therefore, magnetic topological materials have emerged as an ideal platform for relating the transport behaviour and electronic structures in the external fields.

It is known that the electronic structures can be significantly influenced by the spin configuration by breaking or preserving the mirror symmetry [14]. Therefore, modulating the spin structure using a magnetic field is a potential strategy for realising the modulation of topological bands. According to the existing theories and experiments, two main effects of the magnetic field modulating the topological states are the movement of Weyl nodes and the generation of topological states. Theoretical calculations have predicted that the position of the Weyl nodes could be manipulated through magnetisation rotation and the Zeeman splitting during the magnetisation process, resulting in a varying AHE [15-17]. In addition, the mixing of electronic bands with equal or opposite spin owing to spin-orbit coupling was proposed based on the spin-canting effect, resulting in a nonlinear geometrical Hall effect [18]. However, there is no solid experimental evidence for the studies mentioned above, and the related physical mechanism is still being developed. Therefore, further studies are essential.

$EuB_6$, which has a simple band structure (see Supplemental Material, Sec. I [19]), was recently predicted to be a magnetic Weyl semimetal [26,29,30]. The magnetization measurements showed two magnetic transitions at 15.3 and 12.5 K, corresponding to the magnetic moments along the [100] and [111] directions, respectively [31]. The theoretically calculated magnetocrystalline anisotropy energy of $EuB_6$ was only 8 meV [26], which indicated that the orientation of the magnetic moments could be easily driven by a magnetic field. In this case, the magnetic field controlling the spin orientation in real space changed the mirror symmetry of the system and affected the topological energy structures in momentum space. Therefore, the magnetic Weyl semimetal $EuB_6$ is a potentially ideal system for modulating the topological bands using a magnetic field.

In this Letter, by combining transport measurements, theoretical calculations, and scanning tunnelling microscopy/spectroscopy (STM/S), we present strong evidence that the magnetic field can significantly modulate the topological bands and further produce unconventional transport behaviour through the field-induced spin-canting effect in magnetic Weyl semimetal $EuB_6$.

$EuB_6$ single crystals were prepared using the flux method [32] and were confirmed by X-ray diffraction (see Supplemental Material, Sec. II [19]). $EuB_6$ crystallises in a $CaB_6$ structure with a space group of 221 (Fig. 1a). Two significant magnetic transition temperatures $T_{C1}$ and $T_{C2}$ can be observed in inset of Fig. 1b, which indicates that the spin orientation changes from [100] to [111] direction as the temperature decreases, and that the easy axis is in the [111] direction [31] (see also Fig. 1a). Figure 1c shows the magnetic field dependence of isothermal $M$ (upper panels) and Hall resistivity $\rho_{yx}$ (lower panels) at 2 K. The magnetisation curves for $H$ along the [100], [110], and [111] directions exhibited weak magnetocrystalline anisotropy. The dashed lines shown in Fig. 1c (lower panels) represent the linear fit of $\rho_{yx}$ above saturation field $H_S$. In the case of $H$ // [100] and [111], the dashed lines did not pass through the original point at 2 K, indicating the existence of the conventional anomalous Hall effect (CAHE). However, in the case of $H$ // [110] the dashed line did pass through the original point, meaning that the CAHE is unobvious. Interestingly, an additional component of AHE was unexpectedly observed before $H_S$ for $H$ // [100] and [110]. A significant Hall resistivity peak, referred to UAHE ($\rho_{yx}^{UA}$), emerged before the magnetisation saturation of the system. With increasing temperature, $\rho_{yx}^{UA}$ became weaker and vanished, and only the CAHE remained (see Supplemental Material, Sec. III [19]). Notably, a very small UAHE for $H$ // [111] was attributed to the inevitable direction tilt during the cutting of the single crystals.

Based on $\rho_{yx}^{UA} = \rho_{yx} - \rho_{yx}^{N} - R_S M$, where $R_S$ can be determined using the data above $H_S$ by considering $\rho_{yx}^{UA} = 0$ and $M = M_s$ after magnetisation saturation, the $\rho_{yx}^{UA}$ can be extracted (see Supplemental Material, Sec. IV [19]). Furthermore, the unconventional anomalous Hall conductivity ($\sigma_{xy}^{UA}$) can be calculated by the relation $\sigma_{xy}^{UA} = \rho_{yx}^{UA} \big/ (\rho_{yx}^{UA2} + \rho_{xx}^{2})$. As shown in Fig. 1d, for $H$ along [100], [110], and [111], the maximum $\sigma_{xy}^{UA}$ values were 280, 976, and 120 $\Omega^{-1}$ $cm^{-1}$, respectively. In addition, the unconventional anomalous Hall angle (UAHA) was obtained using the relation $\mathrm{UAHA} = \rho_{\mathrm{yx}}^{\mathrm{UA}} \big/ \rho_{\mathrm{xx}}$, as shown in Fig. 1e. The maximum UAHA values were 4.4, 9.6, and 1% for [100], [110], and [111], respectively. UAHE in $EuB_6$ is much larger than CAHE, and is also larger than that for many FM systems [33-39].

The UAHE is very similar to the topological Hall effect (THE) in materials with topological spin textures in real space [39-42], in which Dzyaloshinskii–Moriya interaction (DMI) plays an important role. However, it is considered that DMI does not exist in centrosymmetric $EuB_6$. Therefore, the UAHE in $EuB_6$ most likely results from the evolution of the Berry curvature in momentum space. A similar electronic structure analysis was also performed in the magnetic Weyl semimetal $EuCd_2As_2$ [43]. For $EuB_6$, a very low magnetocrystalline anisotropy energy of 8 meV was reported in a previous theoretical study [26]. This indicates that the competition between the magnetocrystalline anisotropy energy and the Zeeman energy gained in the magnetic field balance each other. In this case, the applied magnetic field potentially induced the noncollinear spin canting, which may further alter the topological bands and electronic transport properties of $EuB_6$. The observation that the $\rho_{yx}^{UA}$ peak was weak when the field was along the [111] easy axis and intense when the field was along along [100] and [110] indicated that a possible spin-canting behaviour ensues when the external magnetic field deviates from the easy axis.

Based on the deduction of spin-canting from the transport results, we performed first-principles calculations to examine the effect of non-collinear magnetism on the electronic structures (see Supplemental Material, Sec. I [19]). Starting from the *z*-directed FM order, we added in-plane spin components ($m_{//}$) to different Eu atoms (marked in Fig. 2c), thereby doubling the size of unit cells along all three lattice vectors. The energy difference was 8 meV when the canting angle θ was 45° ($\tan\theta = m_{//}/m_z$), which indicates that the spin orientations can be easily tuned using external fields. Because of the band-folding effect, the band inversions (consisting of B-p and Eu-d orbits, see Supplemental Material, Sec. I [19]) appearing at the high-symmetry *k*-points X, Y, and Z folded to the Γ point. When the canted spin components became zero and the system was FM, two-fold degenerate bands existed around the Γ point (marked in Fig. 2a) because of the band-folding owing to the equivalence of the *x* and *y* directions. Upon increasing the canted spin components, the degenerated bands split, and all crossings along Γ-Z path become gapped under SOC, as shown in Figs. 2d and 2e. Although the energy scale of the spin-canting-induced band splitting was small (less than 20 meV at the Γ point), the bands changed considerably upon coupling with the previous single degenerate band (green regions in Figs. 2d and 2e), leading to an enhancement in the Berry curvature in the coupled area. The Berry curvature can be calculated directly using the linear-response Kubo formula [44]. We calculated the Berry curvature $\Omega_{xy}$ along the high-symmetry Γ-Z path, on which $\Omega_{yz}$ and $\Omega_{zx}$ always have the same value because of the $\Theta m_{-110}$ symmetry. Each peak in the Berry curvature spectrum corresponds to the band-crossing regime of the band structure. When the canting angle was zero, the bands arising from the high-symmetry *k*-points X and Y (doubly degenerate bands) did not couple with the bands from the Z point (single degenerate band) (Fig. 2f). Owing to the spin-canting effect, these bands couple and open gaps (Fig. 2h), which induce two additional peaks in the Berry curvature spectrum (left in Fig. 2g). When $H$ was smaller than $H_S$, the electronic band structure could be modulated by this behaviour of the non-collinear magnetism. The folding of the electronic band structure thus resulted in the UAHE during the magnetisation process. However, when $H$ was larger than $H_S$, all moments followed the direction of the external magnetic field, and both the spin-canting effect and UAHE disappeared, and only the CAHE remained owing to collinear ferromagnetism with SOC. Furthermore, when $H$ exceeded $H_S$ at 2 K, the conventional AHC measured in our study was only 24 and 89 $\Omega^{-1}$ $cm^{-1}$ for [100] and [111], respectively (see Supplemental Material, Sec. V [19]), which are consistent with the theoretical calculations for the collinear FM state [26].

When magnetisation re-orientation from easy axis to hard ones in $EuB_6$, the magnetic order of system could take an intermediate nonlinear spin-canting state. This spin-canting state enlarges the magnetic unit cell and lowers the symmetry of system, leading to the folding of the bands and change in local density of states (LDOS), which could be reflected in the variation of d$I$/d$V$ spectra in STM/S measurements. To verify the evolution of the band structure with external magnetic field, we performed low-temperature STM/S studies on the Eu-terminated surface of $EuB_6$ (see Supplemental Material, Sec. VI [19]). As shown in Fig. 3a, the square lattice, which indicates that the surface was cleaved along the (001) crystal plane, was clearly resolved in the high-resolution images. The intensity map of d$I$/d$V$ spectra, which are spatially homogeneous within several nm scale (see Supplemental Material, Sec. VI [19]), shows uniform distribution of LDOS along the red arrow (Fig. 3b). We then applied an external magnetic field perpendicular to the surface and recorded the field-dependent averaged d$I$/d$V$ spectra (Fig. 3c) along the red arrow in Fig. 3a. In this case, the magnetic moments gradually rotated from the [111] easy axis to the [001] axis. Significantly, the line shapes of the spectra show a strong field dependence at small magnetic fields. In $EuB_6$, the saturated magnetic field is about 10 kOe along [001] direction. Thus, the response of d$I$/d$V$ spectra to the magnetic field happens only below 10 kOe. After the magnetic moment saturates (>10 kOe) along [001] direction, no remarkable variation in LDOS is observed upon further increase of the magnetic field. Notably, this field-dependent evolution in the LDOS occurred in a rather broad energy window, suggesting a possible reconstruction of the topological band structures.

We further extracted the field-dependent intensity of the spectra in Fig. 3c, and plotted the intensity values in a series of energy as a function of the magnetic field in Fig. 3d. For all energy cuts, the intensity of the d$I$/d$V$ spectrum showed an initial increase, peaked at approximately 5 kOe, and then gradually decreased as the field reached $H_S$. This observation is reminiscent of the transport measurements (Fig. 1d and 1e), where the UAHE demonstrates a highly concurrent field dependence, with an exactly same maximum at 5 kOe. The spin-canting effect causes the band folding and changes the density of states near the Fermi level (Figs. 2d, e and Figs. 3c, d), which further leads to UAHE. We plotted the intensity map (Fig. 3e) of the negative second derivative of the spectra in Fig. 3c, and the field-dependent STS features could be observed more clearly. The substantial enhancement of negative second derivative during the magnetisation indicates a change in LDOS of $EuB_6$ with the application of external magnetic fields.

To provide a clear physical picture, a schematic of the modulation of the electronic band structure and transverse electronic transport characteristics of the magnetic Weyl semimetal $EuB_6$ during the entire magnetisation process is shown in Fig. 4. According to the magnetic measurement results, the ground-state magnetic moment of $EuB_6$ aligns with the [111] easy axis (Figs. 4a and 4b). For $H$ // [111], there are no spin-canting and band-folding effects during the entire magnetisation process. Therefore, it only shows the CAHE upon the unchanged topological bands, as shown in the upper part of Figs. 4b-e. As long as the moments deviate from the easy axis to the hard axis, the competition between the magnetocrystalline anisotropy and the external magnetic field leads to the spin canting of moments, which results in a band-folding effect in momentum space, leading to splitting of the pristine bands and production of additional bands, as shown by the red colour in the middle and bottom regions of Fig. 4c. With the aid of split bands and SOC-induced anti-crossings, an enhancement of the Berry curvature can be obtained. For $H$ along the [100] and [110] main directions, the enhanced Berry curvature leads to the large UAHE observed during the magnetisation process. Once the system enters the saturation state of magnetisation, both spin-canting and band-folding disappear again, as shown in the middle and bottom regions of Fig. 4d. Therefore, UAHE vanishes, and the CAHE produced by the linear FM state remains, as shown in Fig. 4e. This implies that as long as the magnetic field deviates from the [111] easy axis in $EuB_6$, a dynamic picture of spin-canting, band-folding, and UAHE during the process of magnetisation and demagnetisation is very likely to be observed.

In current studies, the peak of the Hall sign during the magnetisation was widely considered to be THE, attributed to the topological spin structures in real space [39-42]. In our study, we observed the band splitting and LDOS evolution, resulting in a prominent UAHE. An external field can change the magnetic structure and further modulate the topological electronic structures, revealing an interplay between the magnetism and topology. The physical origin of the observed Hall peaks during magnetisation process can also be ascribed to the alteration of topological electronic structure in momentum space, not only the topological spin structure in real space.

In summary, the topological electronic structure was modulated owing to the spin-canting effect in magnetic Weyl semimetal $EuB_6$ during the magnetisation re-orientation with the application of a magnetic field. An emergent UAHE was observed with a maximum UAHC of approximately 1000 $\Omega^{-1}$ $cm^{-1}$ and UAHA of approximately 10%, at a low field of 5 kOe. Combining the experiments and theory, we present strong evidence that the magnetic field significantly modulates the topological electronic state and the unconventional transport behaviour. Our findings provide not only an insight into the relationship between the magnetic and electronic structures but also guidance for manipulating the electronic transport properties through the external magnetic field, particularly in emergent magnetic correlated electronic materials for topological spintronics.

**Acknowledgements**

This work was supported by Fundamental Science Center of the National Natural Science Foundation of China (No. 52088101), the Synergetic Extreme Condition User Facility (SECUF), Beijing Natural Science Foundation (No. Z190009), National Natural Science Foundation of China (Nos. 11974394, 12174426, 12104280), National Key R&D Program of China (No. 2019YFA0704900), the Strategic Priority Research Program (B) of the Chinese Academy of Sciences (CAS) (XDB33000000), the Key Research Program of CAS (No. ZDRW-CN-2021-3), CAS Project for Young Scientists in Basic Research (YSBR-003), and the Scientific Instrument Developing Project of CAS (No. ZDKYYQ20210003).

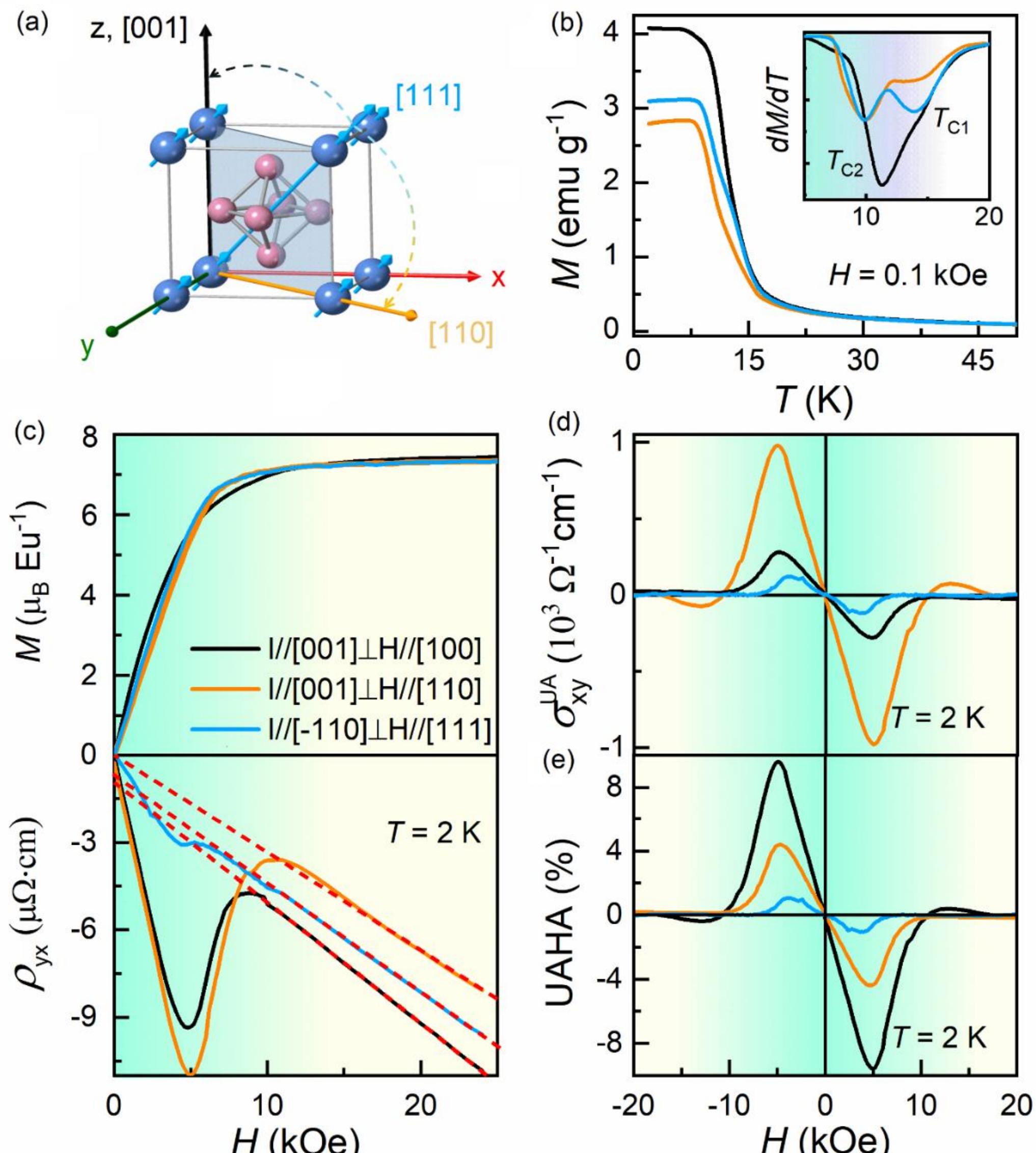

FIG. 1. (a) Crystal structure of $EuB_6$. Blue arrow indicates that the magnetic moment is along the [111] direction at ground state. (b) Temperature dependence of *M* with the field cooling model for *H* = 0.1 kOe along [100], [110], and [111]. Inset shows the temperature dependence of *dM*/*dT*. (c) Magnetic field dependence of isothermal *M* (upper panel) and Hall resistivity $\rho_{yx}$ (lower panel) at 2 K for *H* along [100], [110], and [111], respectively. (d) and (**e**) Unconventional anomalous Hall conductivity ($\sigma_{xy}^{UA}$) and unconventional anomalous Hall angle (UAHA) at 2 K as a function of the magnetic field for *H* along [100], [110], and [111], respectively.

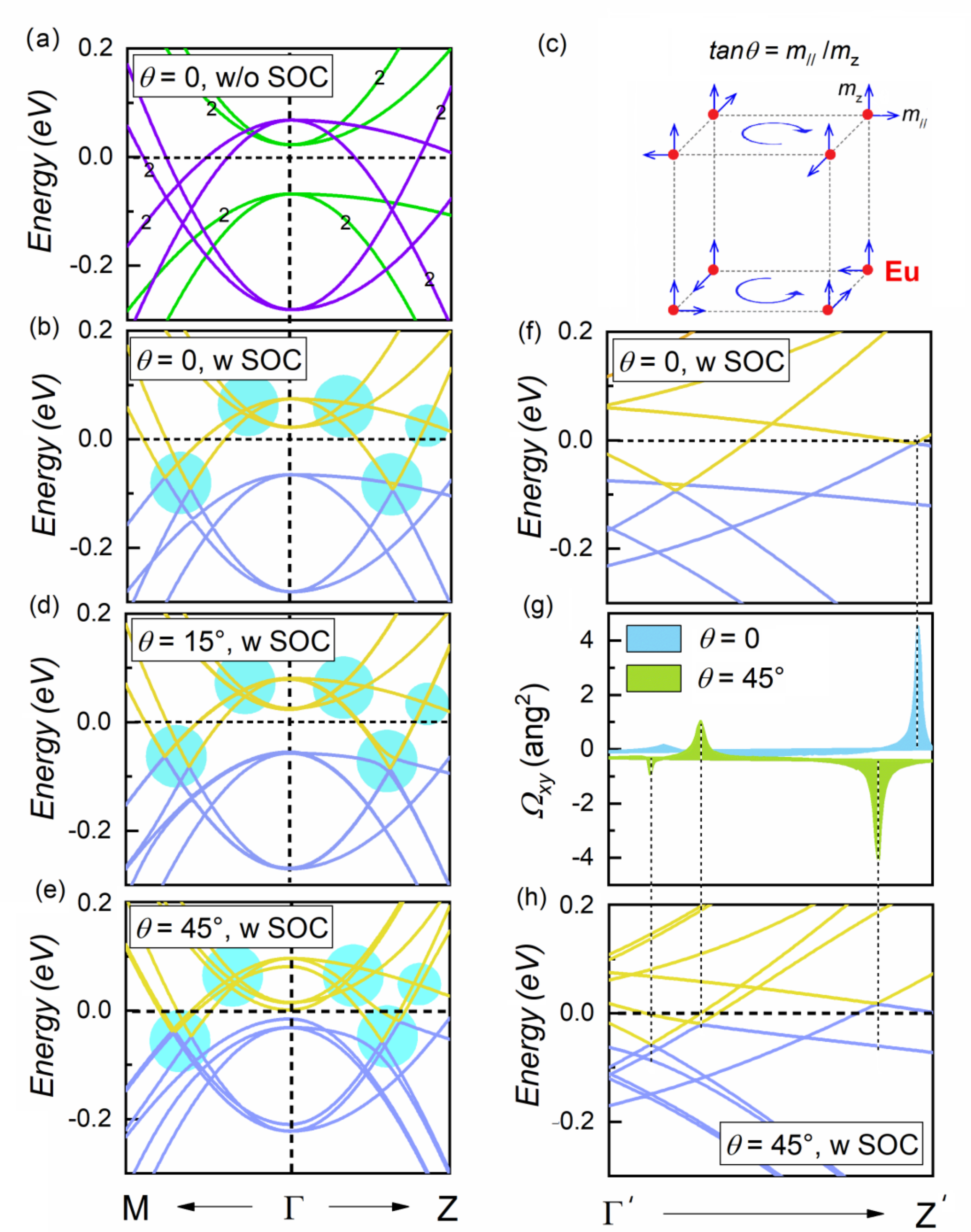

FIG. 2 (a) FM-order band structure for enlarged supercell without considering SOC. Purple and green lines represent spin-up and spin-down bands, respectively. Doubly degenerate bands are marked specifically. (b) FM-order bands after considering SOC. Yellow and light purple lines represent conduction and valence bands, respectively. (c) Schematic of magnetic orientation on each Eu atom in the supercell. Spin-canting band structures with different in-plane magnetic moments for canting angle θ = 15° (d) and 45° (e). Band-crossing regime for θ = 0° (f) and 45° (h) along the Γ-Z path. (g) Calculated Berry curvature $\Omega_{xy}$ around the band-crossing regime along the Γ-Z path.

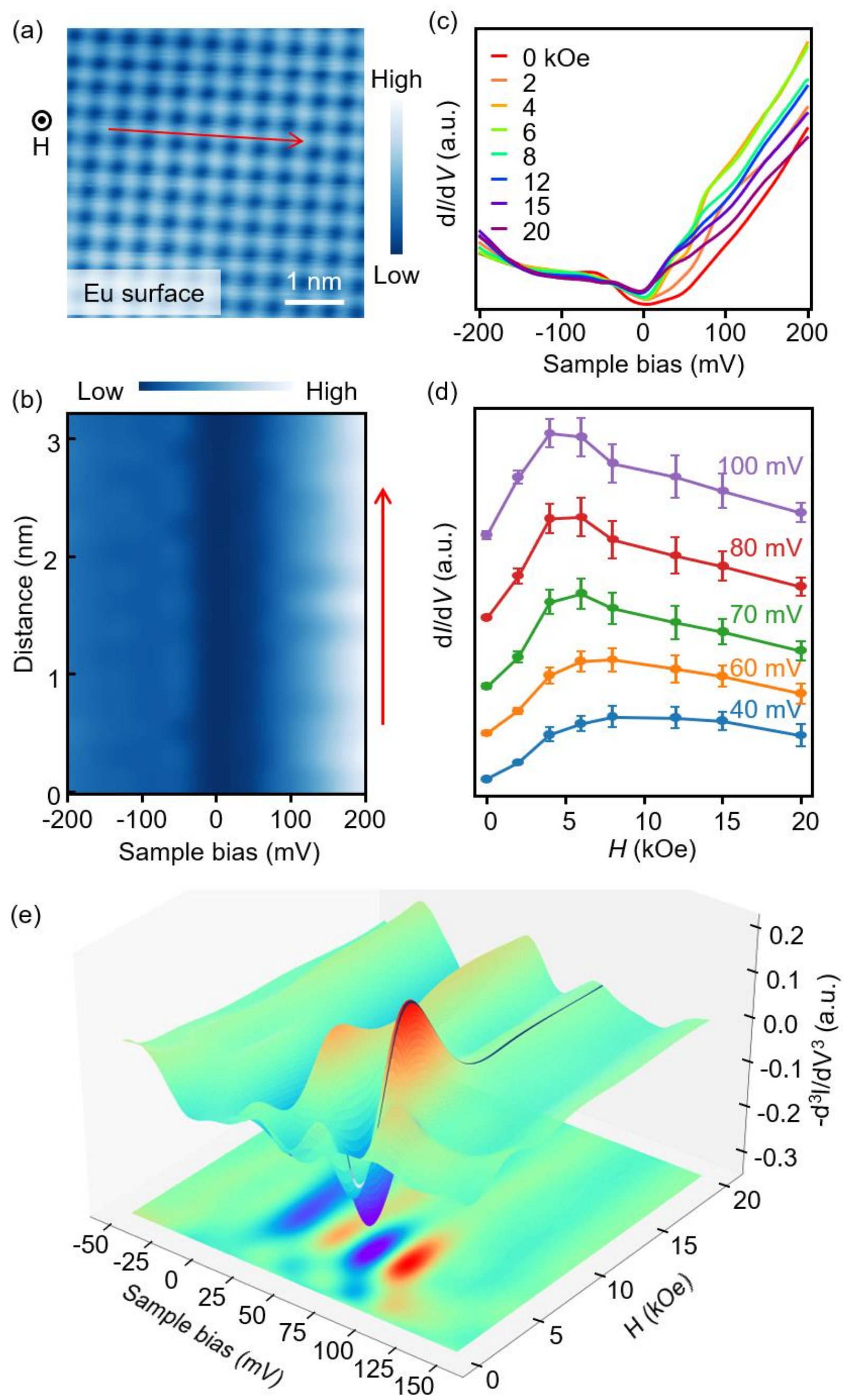

FIG. 3 (a) Atomically resolved image of the Eu-terminated (001) surface, showing the square Eu lattice. The applied magnetic field is perpendicular to (001) surface. Scanning settings for (a): bias $V_s$ = -200 mV, setpoint $I_t$ = 200 pA. (b) Intensity map of fifteen d$I$/d$V$ spectra taken along the red arrow in (a) under zero magnetic field. (c) Field-dependent d$I$/d$V$ spectra taken under magnetic fields from 0 to 20 kOe. Each spectrum is the average of fifteen spectra taken along the red arrow in (a), but under different magnetic fields. Setpoint current of each spectrum is kept the same at 200 pA. (d) Intensity plot of the spectra in (c) under different energies as a function of the magnetic fields. The errors come from the standard deviations of fifteen different spectra taken along the red arrow in (a). (e) Intensity plot of the negative second derivative of the spectra in (c). All the STM/S measurements are taken under 0.45 K.

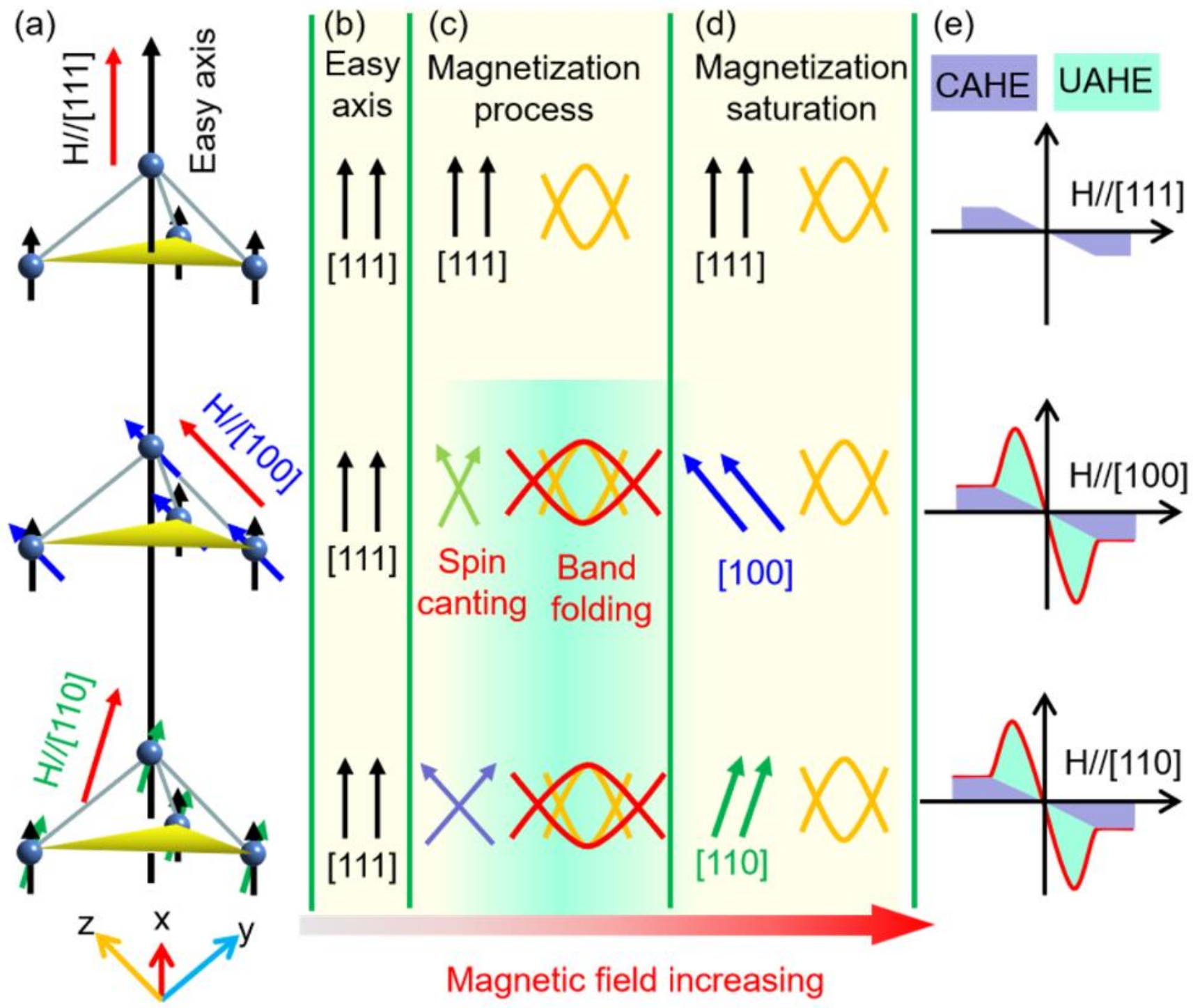

FIG. 4 (a) Magnetic fields along [111], [100], and [110] directions and the rotations of the magnetic moments. (b) Magnetic moments along [111] easy axis in the ground state below $T_{C2}$. (c) Spin-canting and band-folding effects in the low-field magnetisation process for *H* along [100] and [110]. There are no spin-canting and band-folding effects for *H* along [111] easy axis. (d) Magnetic moments along the external magnetic field and corresponding electronic bands for saturated magnetisation. Spin-canting and band-folding effects disappear again in this case. (e) CAHE and UAHE for *H* along different crystal axes. Red arrow at the bottom indicates the increasing magnetic field along the main crystal axes.